\documentclass[10pt]{article} 
\usepackage{booktabs} 
\usepackage{graphicx}
\usepackage{tabularx}
\usepackage{color}
\usepackage{hyperref}
\usepackage{xargs}
\usepackage{algorithm}
\usepackage[noend]{algpseudocode}
\usepackage{listings}
\usepackage{xcolor}
\usepackage{multirow}
\usepackage{colortbl}
\usepackage{hhline}
\usepackage{xspace}
\usepackage{array}
\usepackage[outline]{contour}
\usepackage{blindtext}
\usepackage{appendix}
\usepackage{subcaption}
\usepackage{arydshln}
\usepackage{float}
\usepackage{makecell}
\contourlength{0.3em}
\contournumber{250}

\usepackage[preprint]{tmlr}

\usepackage{amsmath,amsfonts,bm}

\def\eqref#1{equation~\ref{#1}}

\def\1{\bm{1}}

\DeclareMathAlphabet{\mathsfit}{\encodingdefault}{\sfdefault}{m}{sl}
\SetMathAlphabet{\mathsfit}{bold}{\encodingdefault}{\sfdefault}{bx}{n}

\usepackage{hyperref}
\usepackage{url}
\newcommand{\slicedkex}{\emph{SlicedKex}}
\newcommand{\smartkex}{\emph{SmartKex}}
\newcommand{\metakex}{\emph{MetaKex}}
\newcommand{\headerkex}{\emph{HeaderKex}}
\newcommand{\graphkex}{\emph{GraphKex}}

\title{Bridging the Semantic Gap in Virtual Machine Introspection and Forensic Memory Analysis}

\author{\name Christofer Fellicious \email christofer.fellicious@uni-passau.de \\
      \addr Chair of Data Science\\
      University of Passau
      \AND
      \name Hans P. Reiser \email hansr@ru.is \\
      Reykjavik University
      \AND
      \name Michael Granitzer \email michael.granitzer@uni-passau.de \\
      \addr Chair of Data Science\\
      University of Passau
      }

\def\month{04}  
\def\year{2025} 
\def\openreview{\url{https://openreview.net/forum?id=XXXX}} 

\begin{document}
\maketitle

\begin{abstract}
Virtual Machines~(VM)~are becoming increasingly popular and Forensic Memory Analysis~(FMA)~and~Virtual Machine Introspection~(VMI)~are critical tools for security in a virtualization-based approach.
VMI and Forensic Memory Analysis~(FMA)~involves using digital forensic methods to extract information from the system to identify and explain security incidents.
Furthermore, VMI operates on live systems, analysing active processes and detecting threats in real-time, whereas FMA follows a more static approach by extracting and analyzing memory snapshots after an incident has occurred.
A key challenge in both FMA and VMI is the \textbf{``Semantic Gap"}, which is the difficulty of interpreting raw memory data without specialized tools and expertise.
In this work, we investigate how a priori knowledge, metadata and engineered features can aid VMI and FMA, leveraging machine learning to automate information extraction and reduce the workload of forensic investigators.
We choose OpenSSH as our use case to test different methods to extract high level structures.
OpenSSH is a widely adopted protocol for secure server-client communication.
We also test our method on complete physical memory dumps to showcase the effectiveness of the engineered features.
Our features range from basic statistical features to advanced graph-based representations using malloc headers and pointer translations.
The training and testing are carried out on public datasets that we compare against already recognized baseline methods.
We show that using metadata, we can improve the performance of the algorithm when there is very little training data and also quantify how having more data results in better generalization performance.
The final contribution is an open dataset of physical memory dumps, totalling more than 1 TB of different memory state, software environments, main memory capacities and operating system versions.
Our methods show that having more metadata boosts performance with all methods obtaining an F1-Score of over 80\%.
Our research underscores the possibility of using feature engineering and machine learning techniques to bridge the semantic gap.
Our source code and datasets are open source and available online\footnote{\url{https://anonymous.4open.science/r/MachineKex-78F3/README.md}}.
\end{abstract}






\section{Introduction}
\label{machinekex_sec:introduction}
Virtual machines~(VMs) are gaining traction in the tech world, not only for their efficiency in resource utilization but also for their versatility in adapting to various computing environments.
One of the key advantages of VMs is their inherent isolation, which prevents the guest operating system from interfering with the hypervisor's control over the virtual environments.
This isolation allows multiple VMs to reside sandboxed on the same hardware to share resources efficiently without affecting each other.
The extensive usage of VMs also has certain downsides regarding security, and the number of attacks increases yearly.
Checkpoint, a leading cybersecurity platform, reports that attacks increase yearly with 2023 seeing an almost 38\% over the year 2022\footnote{\url{https://blog.checkpoint.com/2023/01/05/38-increase-in-2022-global-cyberattacks/}}.
Therefore, having different defence mechanisms to thwart malicious actors is a priority.

Virtual Machine Introspection~(VMI)~and Forensic Memory Analysis~(FMA)~are important security mechanisms for virtualization-based security.
Virtual Machine Introspection (VMI) refers to the process of monitoring and analysing the memory of a running virtual machine in real-time, typically from a hypervisor layer. 
In contrast, Forensic Memory Analysis (FMA) involves the static examination of memory snapshots, typically captured after an event, to investigate and understand the sequence of actions leading to a potential security breach.
A common challenge faced by both methods is the~``\textit{semantic gap}", ~which arises when developers attempt to interpret raw memory data without high-level abstractions typically available within the guest operating system~\cite{dolan2011leveraging}.
This semantic gap complicates the creation of security solutions at the hypervisor level.
When developing software inside an OS, programmers often have rich semantic abstractions such as system calls or APIs~(e.g., getpid)~to inspect kernel states. 
However, there are no such abstractions for guest OS at the hypervisor layer, but rather the zeros and ones of raw memory data.
Consequently, developers must reconstruct the guest OS abstractions from the raw data at the hypervisor layer. 
Another issue, according to Harrison et al., is that using traditional methods to bridge the gap would impact guest applications' performance due to the system's required context switching~\cite{harrison2012constructing}.
Therefore, reconstructing high-level data structures efficiently and effectively is one of VMI and FMA's core focus areas.

In order to defend against malicious actors, we need this information to make security-critical decisions.
This information resides within the guest operating system's primary memory, where we need to reconstruct the high-level structures from the raw memory data.
Traditional approaches to analysing virtual machine memory rely on handcrafted heuristic, static rules, or even manual forensic analysis.
These methods are inherently brittle; they require constant updates as software evolves, often rely on expert intuition, and struggle to scale across different OS versions.
Furthermore, they fail to generalize when attack vectors change, making them ineffective against novel threats.
Given these limitations, an automated, adaptive approach is needed; one that can learn patterns directly from raw memory data wituout requiring constant manual adjustments.
Machine learning offers a compelling solution, which unlike static heuristics, can learn from large datasets, generalize to unseen cases, and automatically identify anomalies without human intervention.
By leveraging machine learning, we can enhance the VMI and FMA domains without incurring significant performance costs.

Although machine learning has been applied to VMI, most solutions seem tailored towards specific applications such as intrusion detection~\cite{melvin2021practicality, mishra2021vmshield}.
We assume this is due to the unstructured data at the VM level. 
The data itself could vary in value due to the different addresses assigned to the memory structures, and even the values within the structures could change.
If we can overcome the roadblocks and use machine learning algorithms, we could improve the domain of both VMI and FMA while reducing the human effort required.
Machine learning can complement human expertise by prioritizing memory regions most likely to contain security-relevant structures, thereby reducing search space and improving forensic efficiency.

Given the complexity of reconstructing OS-level abstractions from raw memory data, we propose using supervised techniques to classify and interpret memory structures automatically.
To evaluate our machine learning-based approach, we apply our methods to OpenSSH process memory dumps. 
OpenSSH provides a structured environment to validate our feature engineering techniques before extending them to full VM memory analysis.
We chose OpenSSH as a use case because it is one of the most widely used software for client-server communication. 
Moreover, it is also open source and standardized, which helps us understand the software's internal memory structures and source code. 
This gives us the advantage of being able to easily test our hypotheses and feature engineering techniques on the OpenSSH process heap dumps rather than whole memory dumps.
We then apply our knowledge from the OpenSSH use case to whole virtual machine memory dumps.

Our contributions are:
\begin{itemize}
    \item a dataset of complete memory dumps of different Operating System versions and memory sizes totalling over 1.5TB when uncompressed.
    \item feature engineering techniques for reconstructing OS-level structures using OpenSSH as a case study.
    \item a machine learning-based method for classifying kernel-level pointers in physical memory dumps, enhancing automated forensic analysis.
\end{itemize}

We organize the paper as follows. 
We explain the terms and the research gap in~\autoref{machinekex_sec:introduction}~and~\autoref{machinekex_sec:related_work}~contains the related work for our work.
We introduce our methods in~\autoref{machinekex_sec:method}~and~all~the results in~\autoref{machinekex_sec:results}. 
The ethical considerations related to our work is present in~\autoref{machinekex_sec:ethical_considerations}, with~\autoref{machinekex_sec:conclusion}~containing the conclusions we draw from our experiments.

\section{Background}
\label{machinekex_sec:background}
Here we describe the background details of the technologies used in our use cases. We explain the software technologies in detail along with the relevant information regarding the data.
As for OpenSSH/SSH, we explain the basic key mechanisms, the data structure that holds the key and how the client and server do the handshaking to generate a common encryption key and we explain all this in~\autoref{machinekex_subssec:ssh}.
We then explain the basic concepts of VMI such as hypervisors and memory snapshots for our extended use-case in~\autoref{machinekex_subsec:vmidumps}.

\subsection{SSH}
\label{machinekex_subssec:ssh}
\begin{figure*}[t]\resizebox{\textwidth}{!}
{\includegraphics{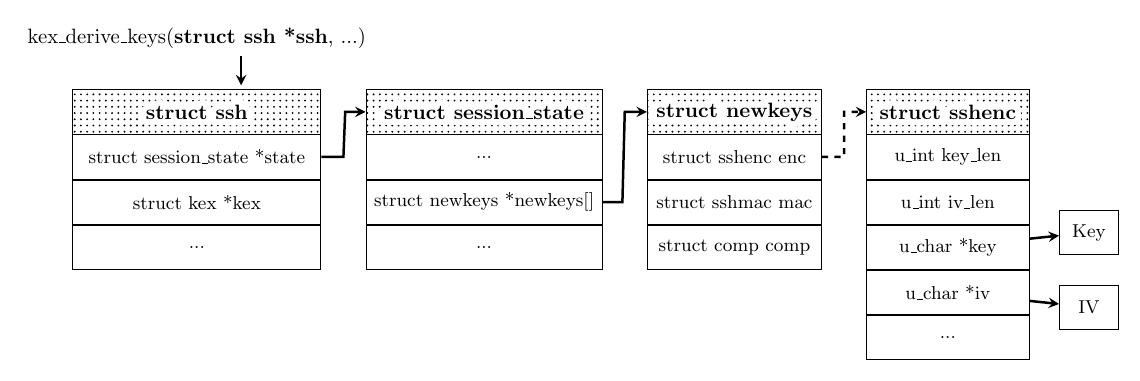}}
\caption{OpenSSH's data structure that holds the encryption key and initialization vector~\cite{sentanoe2022sshkex}. A solid line denotes a pointer that points to the data structure, and the dashed line denotes a direct member of a struct.}
\label{machinekex_fig:ssh_struct}
\end{figure*} 

Secure shell~(SSH) is a protocol to establish a secure remote connection over an insecure network.
The design purpose of SSH was to replace \textit{telnet} and other insecure remote shell protocols.
The client and server use the Diffie-Hellman key exchange to obtain a shared secret.
First, the client generates an ephemeral key pair~(private and public keys) and sends the public key to the server.
Then, the server generates an ephemeral key pair and combines it with the client's public key to generate a master secret $K$ and a hash value $h$.
Next, the server sends the server's public host key $K_S$, its ephemeral public key, and $s$,
a signature of the $h$ calculated with its private host key.
After that, the client validates the $K_S$ using a local database, calculates the $K$ and the $h$, and verifies the signature of $s$ on $h$. 
Finally, the server and the client derive the SSH session keys based on the $K$ and $h$.
The SSH session keys consist of six keys which comprise the Encryption Keys~(EK), Initialization Vectors~(IV), and Integrity Keys~(IK).
\begin{itemize}
     \vspace{-0.3cm}
     \item \textbf{Key A}: IV from client to server
     \vspace{-0.25cm}
     \item \textbf{Key B}: IV from server to client
     \vspace{-0.25cm}
     \item \textbf{Key C}: EK from client to server
     \vspace{-0.25cm}
     \item \textbf{Key D}: EK from server to client
     \vspace{-0.25cm}
     \item \textbf{Key E}: IK from client to server
     \vspace{-0.25cm}
     \item \textbf{Key F}: IK from server to client
     \vspace{-0.25cm}
\end{itemize}
The encryption and integrity check methods dictate the length of each key.
For example,~\textit{aes192-ctr} uses 24 bytes EK and 16 bytes IV.
To decrypt the SSH connection, we can either extract the $K$, $h$, and session ID, then derive the keys ourselves, or extract the EK and IV pair.
For our research, we implement the latter method, as it is more straightforward to extract the session keys rather than doing an additional step for deriving the session keys on our own.

OpenSSH is the most commonly used SSH implementation.
It defines many data structures with different purposes when serving an SSH connection.
One of them is the \texttt{newkeys} data structure.
The \texttt{newkeys} data structure holds the pointers that hold the addresses of the session keys.
Figure~\ref{machinekex_fig:ssh_struct} shows the OpenSSH's data structure that holds the session keys.

\subsection{VMI Dumps}
\label{machinekex_subsec:vmidumps}
\begin{figure}
    \centering
    \includegraphics[width=0.5\linewidth]{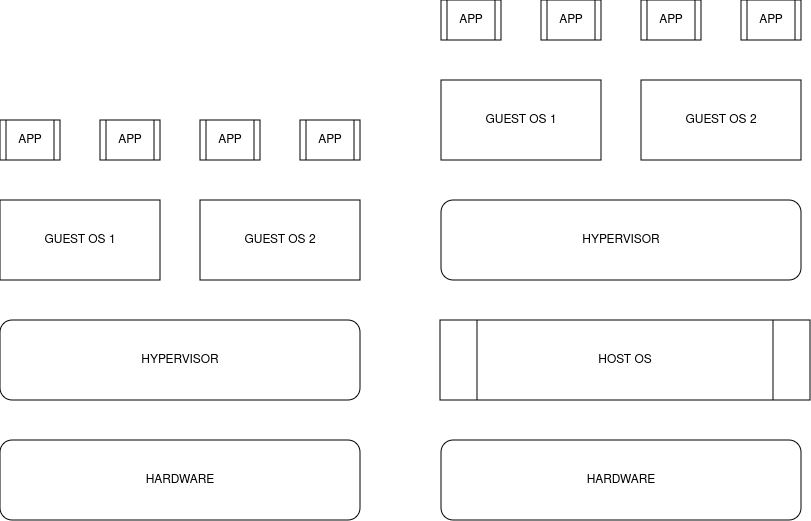}
    \caption{Different types of hypervisors.}
    \label{machinekex_fig:hypervisor}
\end{figure}

In our case, VMI dumps refer to physical memory snapshots taken from a virtual machine. 
We do this using a hypervisor, a software that allows creating and monitoring virtual machines.
Hypervisors are of two types.
\begin{itemize}
    \item Type 1: These run on the hardware directly and are a lightweight operating system. e.g., Zen
    \item Type 2: These hypervisors run within an OS as software and abstract the guest OS~(VMs)~from the host OS. eg: VirtualBox
\end{itemize}
The two types of hypervisors are shown in~\autoref{machinekex_fig:hypervisor}.

The three important properties of a hypervisor are~\cite{garfinkel2003virtual}:
\begin{itemize}
    \item Isolation: Software running in a VM cannot access or modify the software running on the hypervisor
    \item Inspection: The hypervisor has access to the current state of the VM, and this includes the CPU state~(e.g., registers), all memory~(primary and secondary), including its contents, and the state of devices connected to the VM itself.
    \item Interposition: The hypervisor can intervene in the execution flow of the VM if it needs to do so during specific operations, such as executing privileged instructions.
\end{itemize}

The isolation part is important as it guarantees a one-way control flow even if malicious actors take over the VM. 

Taking a memory snapshot involves the interposition property of hypervisors.
We could either pause the VM while writing out the contents of the main memory or do so as needed without pausing.
Pausing the VM has a distinct advantage in that we can be sure the memory is consistent with the addresses and the pages in the memory.

We use Volatility3, a very popular open-source software for VMI applications, to debug symbols and map the virtual memory to the physical memory in the snapshot.
Volatility uses a metadata file that contains information such as the layout and location of the kernel data structures.
Using this metadata file, we could then associate the raw physical data in the snapshot with virtual memory address space.
\subsection{Terminology}
\label{machinekex_subsec:terminology}
We explain the basic terminology used in the paper. It encompasses all the different methods.
\begin{enumerate}
    \item slice: a contiguous block of memory that is of a specific size.
    \item structure: it is a chunk of memory allocated by an allocator such as malloc. A pointer exists within the heap that points to this chunk of memory and the allocated size can be decoded from the malloc header present just before the allocated memory.
    \item pointer\_candidate: an 8-byte block of data that could be a pointer. A pointer candidate is any 8-byte data block that satisfies \autoref{machinekex_eqn:pointer_candidates}. Pointer candidates are searched only in 8-byte aligned memory.
    \item valid pointer: a pointer that points to an address within the heap. This is obtained by translating the address of a pointer\_candidate and making sure that it points to a structure within the current heap and that the allocated size for the structure is greater than zero.
    \item size: the size in bytes allocated to a structure, and the malloc header provides this information.
    \item memory dump/dump: a snapshot of the physical memory of a virtual machine. It contains all the raw data present in the memory at the time of taking the snapshot.
    \item profile: a metadata file that contains information about the layout of the structures, their size and other information. We require the profile to do virtual address translation and extraction of kernel or user structures from the main memory.
    \item virtual address: an address generated by the CPU which is 64-bits long. It does not map 1:1 to the physical memory.
    \item physical address: an address that is actually on the physical memory. There are address translation mechanisms built into the OS for translating virtual addresses to physical addresses.
    
\end{enumerate}

\begin{equation}
    (Address_{base} \leq P_{candidate} \leq Address_{end}) \land (P\%8=0))
    \label{machinekex_eqn:pointer_candidates}
\end{equation}
where, $Address_{base}$ is the starting address of the heap, $Address_{end}$ is the address of the end of the heap and $P_{candidate}$ is a 8-byte memory block converted to a 64-bit unsigned int.

\section{Related Work}
\label{machinekex_sec:related_work}
Dolan-Gavit et al. present a method to address the semantic gap problem in the VMI scenario~\cite{dolan2011leveraging}.
The authors argue for building upon the existing forensic tools for quicker turn around times. 
Using this the authors show that the semantic gap problem could be addressed using existing forensic systems as a partial solution for virtual machine introspection.
Harrison et al. also addresses the semantic gap problem present in VMI and FMA applications~\cite{harrison2012constructing}.
The authors "implement a production oriented prototype utilizing designs that combines and narrows the semantic gap in a modular framework".

Purnaye and Kulkarni created memory dumps of virtual machines containing around 360 Virtual Machine (VM) dumps with a total dataset file size of approximately 80GB zipped\cite{prasad2020}. This dataset contains dumps of continuously generated data. 
The Dumpware10 dataset covers 4294 samples from 10 different malware families~\cite{bozkir2021catch}. This dataset contains 3686 malware and 608 benign samples.
The dataset can be represented as RGB images and has the advantage that computer vision methods are compatible with the dataset. Sadek et al. created a dataset compromising Windows 10 VMs~\cite{sadek2019memory}. The authors deliberately infected Windows 10 VMs and collected ten snapshots of the VM once the malicious payload was running. The snapshots are in the "Advanced Forensic Format" (AFF4), which is a compressed format. The compressed file size is around 1GB per snapshot. The dataset contains 1530 snapshots.

Petrik et al. developed a method to analyze raw binary data extracted from the memory dump of a device~\cite{petrik2018towards}. The authors use machine learning methods and a multi-hundred Terabyte dataset to detect malware in memory dumps with a very high success rate. This model aims to be architecture and Operating System (OS) independent for malware detection. Sihwail et al. used a method that combined memory forensics to extract malicious artefacts and generate features for machine learning. The authors report a very high accuracy and low false positive rate. Tran et al. also use memory forensics and machine learning to identify malware data~\cite{tran2021independent}. They implement an OS-independent malware detector which is also geared towards finding unidentified malware.

Taubmann et al. developed a method for extracting the master key of a Transport Layer Security(TLS) connection at runtime\cite{taubmann2016tlskex}. The procedure takes place over two stages: online and offline. 
The execution of the online stage is synchronous with the TLS communication. 
The offline part can be executed at a later point in time when the required decrypted content of the TLS connection is required. The authors also use knowledge of the data structures and pointers within the TLS session objects. The search pattern follows pointers in the memory. The limitation of this method is that the decryption only works on known software and not on unknown software such as malware.

Taubmann et al. developed another method for fast extraction of ephemeral TLS keys\cite{taubmann2018droidkex}. This work targets the Android Operating System. The method extracts the Master Secret(MS) required to derive the symmetric session keys. We need to extract the MS must during a TLS session; otherwise, there is no guarantee that the MS will be present in the main memory. Therefore, the method must be executed synchronously with the application's control flow. The MS is extracted by dereferencing pointers that point to a structure that holds the encryption key material. 

MemDecrypt by McLaren et al. uses a method based on Shannon's theorem of entropy to extract encryption keys\cite{mclaren2019decrypting}. The principle that good encryption keys have to be random bits and thus have a very high entropy forms the basis of this method. The authors use Shannon's measure for discrete entropy for computing entropy\cite{shannon2001mathematical}. The authors "focus on SSH decryption using AES-CTR and AES-CBC in virtualized environments using MemDecrypt." The method defines thresholds for different sizes of memory segments. Furthermore, decryption attempts are done on the network traffic only on memory segments above the threshold. A potential downside of this method is that we need to identify appropriate thresholds and fine-tune them manually for all potential key lengths. 

Sentanoe and Reiser developed a method named "SSHKex" for extracting OpenSSH keys from VMI\cite{sentanoe2022sshkex}. This method differed from the previous method in that it did not require any modifications to the server and only paused the Virtual Machine twice to extract the session keys for each session. SSHKex uses the debugging symbol information to extract the SSH keys. For decryption, the procedure uses two decryption tools. The first is a Python library, which is used as a proof-of-concept to test whether the correct keys are retrieved. The second tool is a Wireshark plugin. \textit{Wireshark} is an open-source packet sniffing software\cite{combs2008wireshark}.  The Wireshark plugin will decrypt and print the network communication plain text when the user inputs the SSH session keys. The authors claim that their method increases the connection time by only 50mS on average and under heavy load on the server, an overhead of 85.8ms.

Vergeer used a similar method to that of Sentanoe and Reiser to extract the SSH session keys\cite{vergeer_2020}. The author also uses the specific properties of the data structures of OpenSSH to extract the keys. The method scrapes the entire memory address space, and each offset is validated against certain constraints. Finally, the method uses a network parser to decrypt SSH sessions into a readable output. The scripts and plugins for the method are available.

Fellicious et al. published the OpenSSH dataset and developed a method that uses Random Forest Classifiers to identify windows of data that contain keys\cite{fellicious2022smartkex}. Their method involved slicing the data into 128-byte chunks based on the entropy of the data. Data parts with repeated patterns are discarded as there is little to no chance of it being an encryption key. The authors also use Synthetic Minority Over-sampling Technique(SMOTE) to increase the recall by sacrificing precision. The next step is brute force the positively identified results to extract the encryption keys from the 128-byte slice. This method is highly effective, with a recall of $99.62\%$ for the oversampled dataset. In addition, the method is fast, with the average execution time for extracting keys from one heap being approximately 0.4s. The downside of this method is that it requires a large amount of main memory to train the model. Another downside of this method is that it requires the final key to be extracted by the brute force method. 
\section{Method}
\label{machinekex_sec:method}
This section presents our approach to reconstructing OS-level memory structures from raw memory data using machine learning techniques. 
Our methodology aims to bridge the semantic gap in VMI and FMA by leveraging metadata to infer and classify memory structures efficiently. 
We evaluate how varying amounts of a priori knowledge influence the accuracy and reliability of memory structure reconstruction across different scenarios, including full virtual machine memory dumps and process-specific data from OpenSSH.

We begin by establishing a baseline method~(\slicedkex), which does not use metadata and instead applies a simple chunking strategy to segment raw memory data. 
This method provides a reference point to assess how feature engineering improves classification performance. We then incrementally incorporate metadata:~(\metakex)~adds ASCII character distributions and zero-byte counts,~(\headerkex)~integrates memory allocation headers,~and~(\graphkex)~further enhances structure identification by modelling memory as a graph of pointers and relationships.
While OpenSSH serves as an evaluation case study, we extend our findings to full virtual machine memory dumps, applying the learned techniques to classify and reconstruct OS-level structures beyond just SSH-related memory regions.
Our code is available online in an anonymized repository at~\url{https://anonymous.4open.science/r/MachineKex-78F3/README.md}.

\subsection{Dataset}
\label{machinekex_subsec:dataset}
To evaluate the effectiveness of our methods, we utilize both an existing dataset and a newly created dataset. 
For process-specific analysis, we use a publicly available dataset from~\cite{fellicious2022smartkex}, which contains heap dumps of OpenSSH processes collected under various execution scenarios, including the base OpenSSH version, SCP transfers, client-side operations, and port-forwarding. 
This dataset offers several advantages:
\begin{itemize}
\item It provides separate training and validation sets, enabling systematic evaluation.
\item The virtual addresses are already translated, simplifying memory structure analysis.
\item It includes established baselines using both machine learning and non-machine learning methods.
\end{itemize}
These properties make it well-suited for assessing different memory structure classification techniques while ensuring direct comparability with prior work.

For full virtual machine memory dumps, no publicly available standardized dataset exists. 
Therefore, we construct our own dataset, spanning multiple OS versions and memory configurations, to evaluate forensic memory analysis techniques more comprehensively. 
The dataset, totalling over 1.5TB~(uncompressed), will be publicly released upon acceptance of this paper.
\subsection{Without any metadata information}
\label{machinekex_subsec:slicedkex}
The most straightforward machine learning-based baseline for SSH key extraction is to chunk the dataset into specific blocks of overlapping sizes. 
We use a chunk size of 128 bytes~(twice the largest key size of 64 bytes)~with a 64-byte overlap.
This 64-byte overlap is a crucial aspect of the process, as it ensures no keys are missed during the chunking process, given that the maximum key size in the dataset is 64 bytes. 
This overlap effectively increases the size of the training dataset by 50\%, though we mitigate this increase somewhat by discarding blocks that contain only zeroes. 
These overlapping chunks form the basis for training our model, providing a comprehensive set of data points for training.

This baseline serves as the foundation for our~\slicedkex~algorithm, described in \autoref{machinekex_alg:slicedkex}. The extracted features from each heap are concatenated to form the final dataset for training and evaluation.

\begin{algorithm}
  \caption{Creating the feature vector without any feature engineering.}
  \begin{algorithmic}[1]
    \Procedure{CreateSlicedKexDataset}{$heapDump, Keys$}
        \State $features \gets \emptyset$
        \State $labels \gets \emptyset$
        \State $length \gets length~of~heap$
        \State $idx \gets 0$
        \State $size \gets size~of~a~chunk$
        \State $overlap \gets 64$
        \While{$idx \leq length~-~size$}
            \State $chunk \gets i = 1...size~;~heapDump(i)$
            \If{$sum(chunk)~=~0$}
                \State $Discard~chunk$ \Comment{Discard chunks with only zeros}
                \State $idx \gets idx~+~overlap$
            \EndIf
            \If{$\exists Keys \in chunk$}
                \State $label \gets 1$
            \Else
                \State $label \gets 0$
            \EndIf
            \State $features \gets features \cup chunk$
            \State $labels \gets labels \cup label$
        \EndWhile
    \EndProcedure
  \end{algorithmic}
  \label{machinekex_alg:slicedkex}
\end{algorithm}

\subsection{Adding engineered features}
In \autoref{machinekex_subsec:slicedkex}, the algorithm operates directly on raw memory data without any preprocessing or feature engineering. 
However, encryption keys have distinct properties—they are high-entropy, randomly structured sequences designed to resist pattern-based attacks. 
This suggests that additional metadata could help the classifier distinguish OpenSSH encryption keys from other memory content. 
To leverage this, we introduce feature engineering techniques that enrich the dataset with informative characteristics. 
Instead of extracting features at the chunk level, which risks obscuring key-specific patterns, we compute features at an 8-byte word level. 
This choice balances granularity and feature vector size, as 8 bytes align with the word size of modern 64-bit architectures. 
By refining the representation in this way, we aim to enhance the classifier's ability to identify encryption keys effectively.
We settled on three simple features for~\metakex, and they are:
\begin{itemize}
    \item Sum of the 8-byte word: We chose this feature because encryption keys exhibit high entropy, resulting in moderately high values. Extremely high or low values suggest non-key words, helping the classifier discard irrelevant chunks.
    \item The number of non-zero values: Some words have a specific number of zeros, which could help identify the non-key words. Encryption keys have high entropy and this value would be low for chunks having high entropy.
    \item The number of ASCII characters: This feature will help us discard random strings such as email or IP addresses which are inherently random but cannot be encryption keys.
\end{itemize}
Most of these features are used to reject the chunks that do not contain any OpenSSH key.
We append these values to the existing feature vector. 
Thus, the feature vector length increases by 48 integer values: 16 per engineered feature, for a total of three features, as shown in \autoref{machinekex_eqn_metakex}.
Therefore, the total feature length is increased to 176.
\begin{equation}
    Total~Feature~Length~=~ChunkSize~+(ChunkSize/WordSize)*3\\
    = 128~+~(128/8)*3=176
\label{machinekex_eqn_metakex}
\end{equation}
The 128-byte feature vector is the same as that created by~\cite{fellicious2022smartkex}.

\subsection{Adding header information}
The OpenSSH implementation is written in C, which uses malloc to allocate runtime data on the process heap, including the SSH encryption keys.
Malloc allocates memory in chunks, each of which is preceded by a malloc header~\cite{delorie_2022}.
We could decode this header information to obtain the size of the allocated chunk and add this to the existing 128-byte slice.
We do this because the encryption keys are of specific sizes, and therefore, the header preceding them should be similar for each size.
We hypothesize that the header information will provide the classifier with additional information, potentially improving performance.

Building on the work of Fellicious et al., who demonstrated that entropy-based methods can be used to reduce the dataset and isolate key regions at the start of each chunk, we leverage this approach to narrow down the key location~\cite{fellicious2022smartkex}.
Fellicious et al. also showed that their method localizes the keys to the start of the chunk.
We use the same method and add the eight bytes before the start of every chunk to include the header information.

\subsection{Graph based extraction}
Sentanoe et al. show that the encryption keys follow a specific structure in the process' heap memory \cite{taubmann2016tlskex,sentanoe2022sshkex}. 
\graphkex~takes it one step further and creates a graph of the pointers, with structures representing nodes and pointers forming the edges between them.
The goal is to represent the heap memory as a directed graph, where nodes correspond to structures allocated using malloc, such as the SSH STRUCT, SESSION\_STATE STRUCT, or NEWKEYS, and edges represent pointers between these structures. 
First, we identify all pointers in the heap, which are 8-byte sequences stored in little-endian format with "00" as the most significant byte~\cite{delorie_2022}.
For each pointer found, we validate the pointer. 
A validated pointer~(valid pointer in this context)~is a pointer that we identified with a malloc header that specifies the allocated space.
Valid pointers are then added as nodes in the graph, where the pointer address is used as the node identifier. 
Invalid pointers are discarded, as it likely represents an invalid pointer, a pointer pointing to the middle of a list~(which has no purpose in our current method) or simply data that looked like a pointer.

After validating pointers, the next step is to search the memory area defined by the malloc header for more pointers. 
The size of the allocated memory block (obtained from the malloc header) defines the search space. 
Each pointer discovered within a structure is checked for validity, and if valid, it is incorporated into the graph by creating a new node or linking it to an existing one. 
Directed edges are added to the graph, connecting parent structures to child structures as defined by the pointers.
A structure in this context is simply any data allocated using~\textit{malloc}, as it can be a simple integer, string, array, structure, or object.
As the traversal of pointers can only take place in one direction~(from the pointer to the memory location), directed edges represent the pointer links in the graph.

The training method starts by looking at all possible pointers within the heap. 
We identify pointers in a heap using the expression in \autoref{machinekex_eqn:pointer_candidates}.

The pointer address is then validated to verify whether the pointer points to within the heap. 
$S_{ptr}$ defines the set of all valid pointers found using the expression~\autoref{machinekex_eqn:pointer_candidates}.
We then create a graph~($G_{heap}$) for the heap dump. 
$G_{heap}$ is a directed graph where the vertices represent identified structures and the edges represent pointers from within one identified structure~(vertex in $G_{heap}$) to the following structure~(another vertex in $G_{heap}$). 
We obtain the memory allocation size for every structure using the \textit{malloc} header information~\cite{delorie_2022}. 
After finding a valid pointer, we check whether it exists in $G_{heap}$; if it does not, we create a node with the node name being the pointer value. 
The next step involves exploring the memory regions defined by the malloc header to identify additional pointers that creates the edges to other nodes and thereby expand the graph structure.
The allocation size of the~\textit{malloc}~header defines the search space. For each pointer identified within this structure, we check if the pointer is valid, and if the pointer is valid, we confirm that the pointer node is present in $G_{heap}$. 
If it is not present, we create a new node with the node name being the pointer address. 
For every identified pointer, we add a directed edge from the parent node~(the node that the pointer points to) to the child nodes discovered within the structure. 
Each node in the graph also has specific properties assigned to it which enables us to easily extract node features for dataset creation.

After parsing the entire heap and constructing the graph, multiple disconnected sub-graphs may be formed. While these sub-graphs represent different regions of memory, we are only interested in the sub-graph that contains the path to the NEWKEYS structure. 
Given the specific structure of the OpenSSH implementation, we know there is a defined pointer path leading from the SSH STRUCT to the encryption keys~\cite{sentanoe2022sshkex}. 
Thus, our focus is on isolating the sub-graph that contains this path, as it will provide the necessary structure to identify and extract the SSH keys.
In order to extract the actual encryption keys, we create a feature vector for every node in the graph.
The feature vector consists of values from the current node and its parent node such as 
\begin{itemize}
    \item size: the size of the current node from the malloc header
    \item pointer count: the number of pointers contained in the structure
    \item last pointer offset: the offset of the last pointer. 
    \item last valid pointer offset: the offset of the last valid pointer. The current and previous properties give us an idea about the structure.
    \item out degree: number of valid pointers in the structure.
    \item parent size: size of the parent node, if present.
    \item parent pointer count: number of pointers in the parent.
    \item offset: offset of the current node pointer in the parent.
    \item parent out degree: number of valid pointers in the parent node.
\end{itemize}
We create a corresponding feature vector for each individual node in the graph and concatenate all feature vectors to form the dataset. 

The algorithm to create the graph and extract the feature vector is in~\autoref{machinekex_alg:graphkex}
\begin{algorithm}
  \caption{Graph Creation}
  \begin{algorithmic}[1]
    \Procedure{CreateGraph}{$heapDump$}
        \State $pointers \gets all~pointers~in~heap$
        \State $G_{heap} \gets emtpy~graph$
        \While{$pointer \in pointers$}
            \If{$pointer is valid$}
                \If {$pointer \notin G_{heap}(v)$}
                    \State $G_{heap}(V) \gets  G_{heap}(V) \cup pointer$
                \EndIf
                \State $size \gets allocation\_size$
                \State $offset \gets 0$
                \State $ptr\_cnt \gets 0$
                \State $valid\_ptrs \gets \emptyset$
                \State $struct\_base\_addr \gets (pointer - base\_address)/8$
                \While{$offset <= size/8$}
                    \State $ptr \gets heapDump(offset+struct\_base\_addr)$
                    \If{$ptr \in pointers$}
                        \State $ptr\_cnt \gets ptr\_cnt + 1$
                        \State $last\_pointer\_offset \gets offset$
                        \If{$ptr~is~valid$}
                            \State $valid\_ptrs \gets valid\_ptrs \cup ptr$
                            \If{$ptr \notin G_{heap}$}
                                \State $G_{heap}(V) \gets G_{heap}(V) \cup ptr$
                            \EndIf
                            \State $G_{heap}(E) \gets G_{heap} \cup (pointer, ptr)$
                            \State $last\_valid\_pointer\_offset \gets offset$
                            \State $offset \gets offset + 8$ 
                        \EndIf
                    \EndIf
                \EndWhile
                \State $G_{heap}(pointer).size \gets size$
                \State $G_{heap}(pointer).pointers \gets ptr\_cnt$
                \State $G_{heap}(pointer).edges \gets count(valid\_ptrs)$
                \State $G_{heap}(pointer).last\_valid\_pointer\_offset \gets count(valid\_ptrs)$
                \State $G_{heap}(pointer).last\_pointer\_offset \gets count(valid\_ptrs)$
            \EndIf
        \EndWhile
    \EndProcedure
  \end{algorithmic}
  \label{machinekex_alg:graphkex}
\end{algorithm}

\subsection{Training and Testing}
We use a random forest classifier with default hyperparameters to ensure a fair baseline comparison with previous results from Fellicious et al.~\cite{fellicious2022smartkex}. 
Random forests are robust to class imbalance and provide a reliable benchmark without the need for fine-tuning.
To evaluate performance, we apply the trained classifier to the entire validation set using a consistent methodology across all methods. 
Instead of training on the full dataset at once, we incrementally increase the training size to observe performance trends as more data is introduced.
We begin with 500 instances, ensuring that all methods are trained on identical subsets to maintain comparability.
Despite controlling for variability in data selection, some fluctuations remain due to the inherent stochasticity of random forests. 
However, this approach allows us to assess how each method performs as the dataset grows while minimizing inconsistencies in training data.

\subsection{Extraction of features on VMI snapshots}

\begin{figure}
    \centering
    \includegraphics[width=0.95\linewidth]{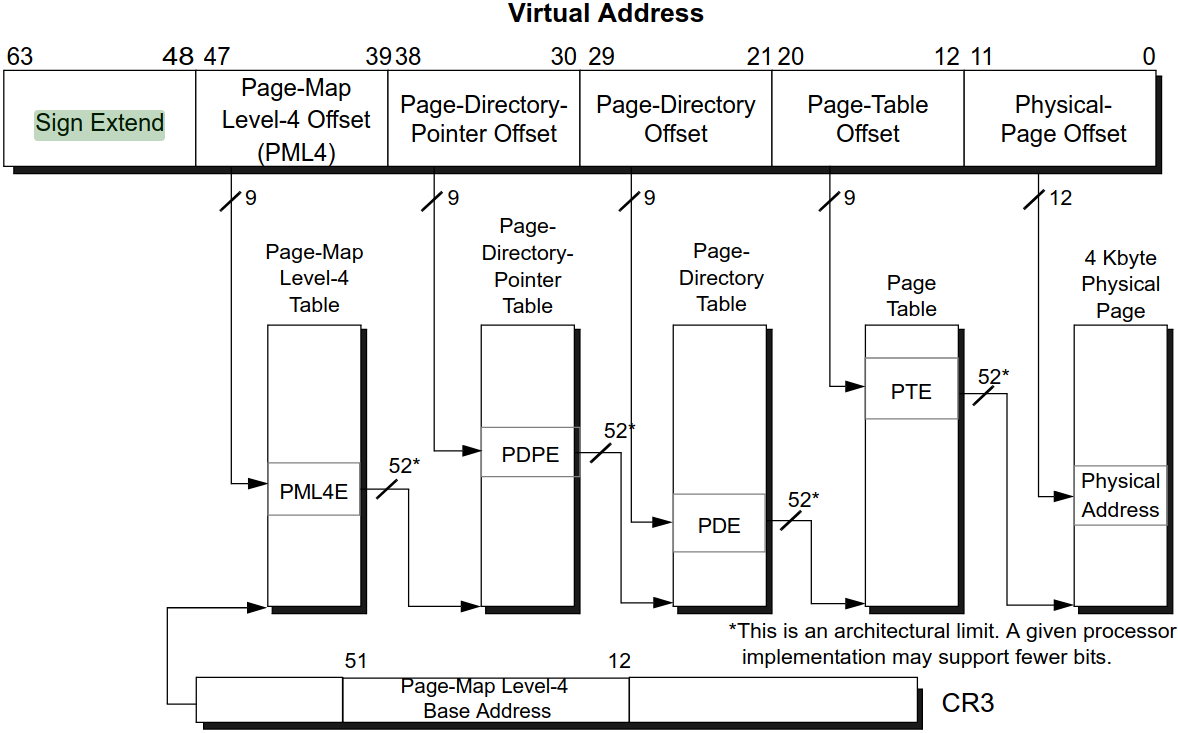}
    \caption{Virtual address to physical address translation for a 4KB page.}
    \label{machinekex_fig:address_translation}
\end{figure}

Using Virtual Machine Introspection~(VMI)~techniques, we can identify system pointers and reconstruct kernel structures from raw memory artifacts.
System pointers in main memory follow a specific pattern, making them easy to identify with a simple regular expression.
We could identify addresses by decoding the 8-byte data and looking at the bits 63:48 of the virtual address as shown in~\autoref{machinekex_fig:address_translation}. 
The sign extension bits mirror bit 47, as required for canonical-address forms\footnote{\url{https://ia600502.us.archive.org/23/items/24593APMV21/24593_APM_v21.pdf}}.
This means that if bit 47 is a 1, all bit from 63:48 will be 1 and vice versa.
For kernel-level pointers, this bit is set to 1, allowing us to identify kernel-space pointers.
We perform virtual-to-physical address translation using the Volatility3 software\footnote{\url{https://github.com/volatilityfoundation/volatility3}}.
Volatility is a well known tool for VMI applications and is open source. 
Volatility3 performs virtual-to-physical address translations and, with the help of custom plugins, extracts memory artifacts such as linked lists and process structures. 
In our approach, we use Volatility3 to resolve virtual addresses and extract regions of interest from physical memory, forming the basis for our metadata-based feature extraction.
We have a simple open source plugin that takes the virtual address and converts it to the appropriate physical addresses, enabling structured memory analysis.

Once we resolve virtual addresses to their corresponding physical locations, we analyse kernel structures that rely on linked lists for data organization. By traversing these linked lists, we extract metadata that characterizes their structure and relationships within memory.
For each pointer that we identify in the physical memory dump, we extract information around the area that the pointer points towards.
We do this because pointers might not always reference the head of a linked list or array and the pointers could reference any node in a linked list.
Also, contiguous locations in virtual memory address space might not be contiguous in the physical address space due to the memory mapping of the virtual address space to physical address space.
For the main memory snapshots, we do not have information about headers to estimate the size of allocation due the pointers not pointing exactly to the head of the structure.
Another reason is that the low level allocation calls in the kernel space differ from the user space based allocation calls\footnote{\url{https://lwn.net/Articles/486301/}}.
Making such an assumption that the pointer always points to the head of the structure in kernel structures is not possible due to the architecture of the Linux Operating System.
Therefore, we craft three different features based on metadata derived from traversing the linked lists which are 
\begin{itemize}
    \item distance: number of steps to a dead end or back to the starting point in a circular linked list.
    \item count: total number of elements in the linked list.
    \item size: estimated structure size, determined by the minimum distance between consecutive structures.
    \item incount: number of incoming references to a target structure.
\end{itemize}

Similar to~\metakex~we add we add these features to the raw data from the memory and thus create the feature vector.
We then train the feature vectors extracted from the training data on random forests.

\section{Results}
\label{machinekex_sec:results}
We compare our method against the SmartKex method from Fellicious et al~\cite{fellicious2022smartkex}.
The SmartKex method reduces the training dataset by looking at areas with high entropy as encryption keys are regions with high entropy.
Additionally, we use SmartKex as the baseline for our other machine learning methods.
Another reason we use SmartKex is due to the training time required for the random forest for~\slicedkex.
\slicedkex~requires more than four days for the whole dataset while~\smartkex~requires only around eight hours.
If we implement~\metakex~on the~\slicedkex~algorithm~it will further increase the training time, therefore, we implement~\metakex~on~\smartkex~to reduce the training times and also use~\smartkex~as the baseline for comparison.
We implement~\headerkex~on the~\smartkex~algorithm as we need the encryption keys to be at the beginning of the chunk and~\smartkex~does exactly that.

We compare the results of all the methods at different training set sizes against a standard validation dataset. 
The validation dataset, published by the authors on Zenodo, is non-overlapping and contains over 15,000 individual files.
We train all our methods using the same seed so that the experiment is deterministic and we test against the same dataset.
This allows us to obtain a better picture on the efficacy of each method and the effects of metadata and engineered features.
The dataset is severely unbalanced, which we expected as there are only a maximum of six encryption keys, with more than 99\% data belonging to the negative class.

\begin{figure}
\centering
\begin{subfigure}{0.45\textwidth}
    \includegraphics[width=\textwidth]{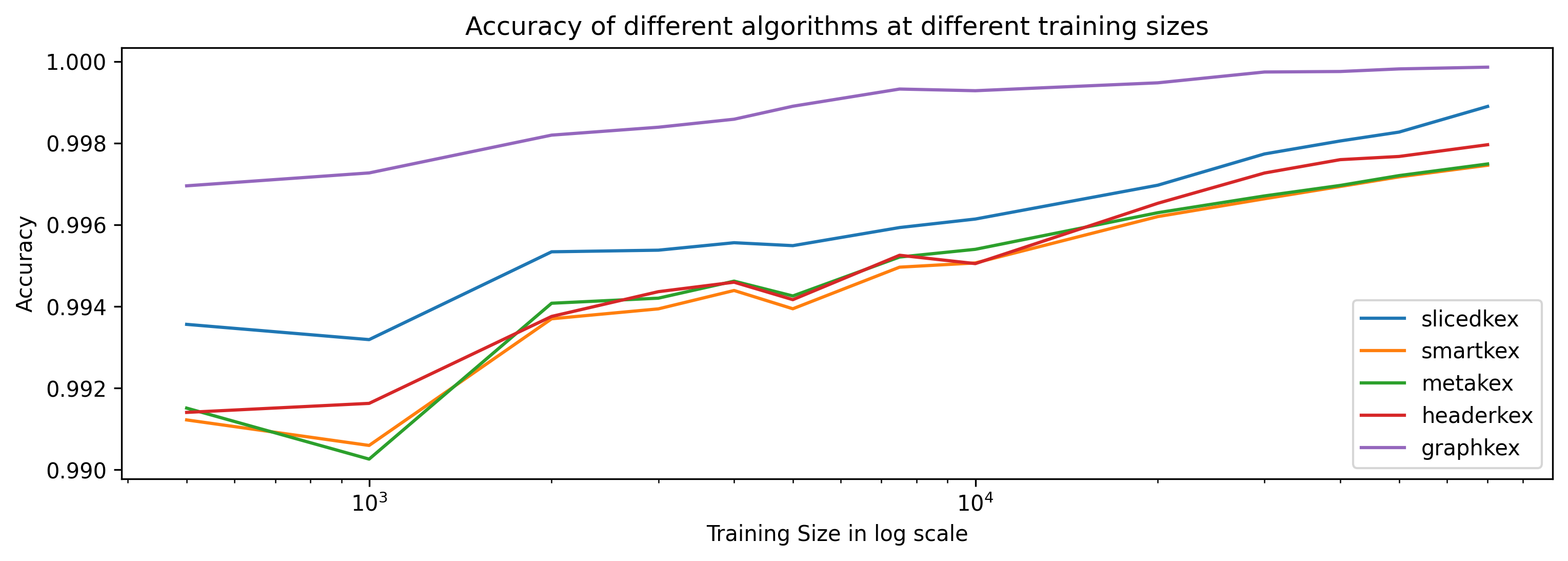}
    \caption{Accuracy of the different methods under different training data sizes. The x-axis is in log scale.}
    \label{machinekex_fig:accuracy}
\end{subfigure}
\hfill
\begin{subfigure}{0.45\textwidth}
    \includegraphics[width=\textwidth]{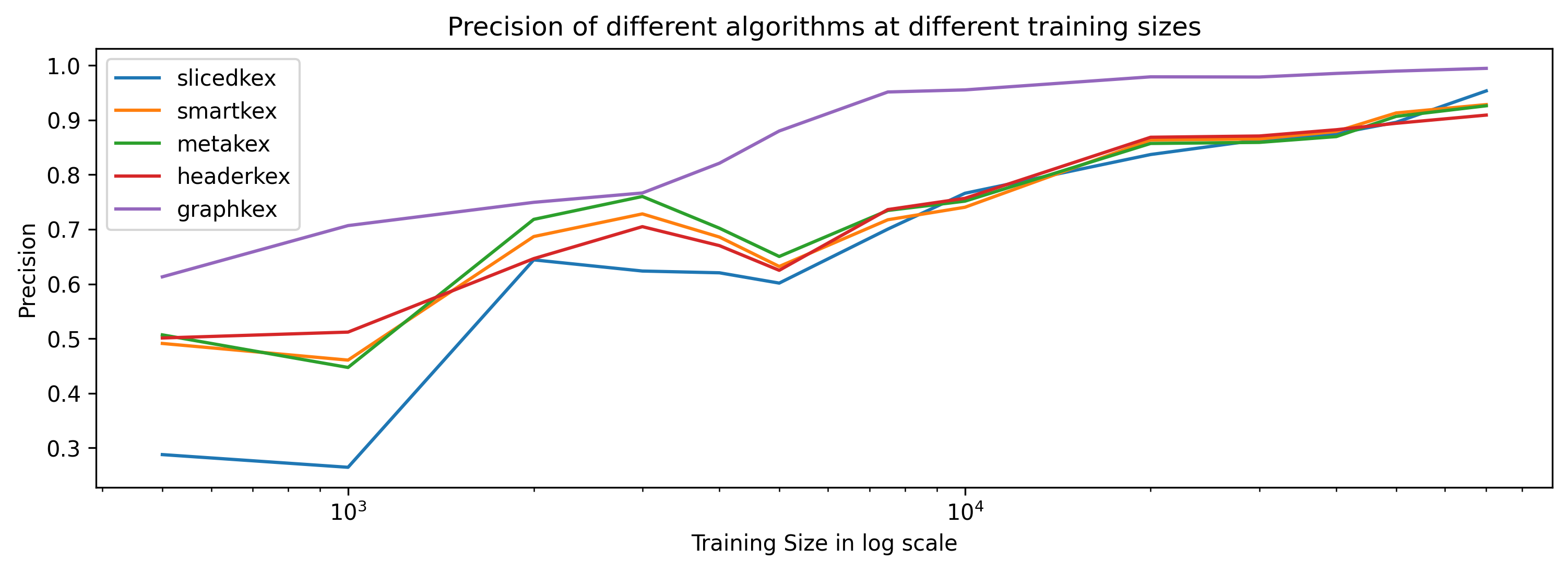}
    \caption{Precision of the different methods under different training data sizes. The x-axis is in log scale.}
    \label{machinekex_fig:precision}
\end{subfigure}
\hfill
\begin{subfigure}{0.45\textwidth}
    \includegraphics[width=\textwidth]{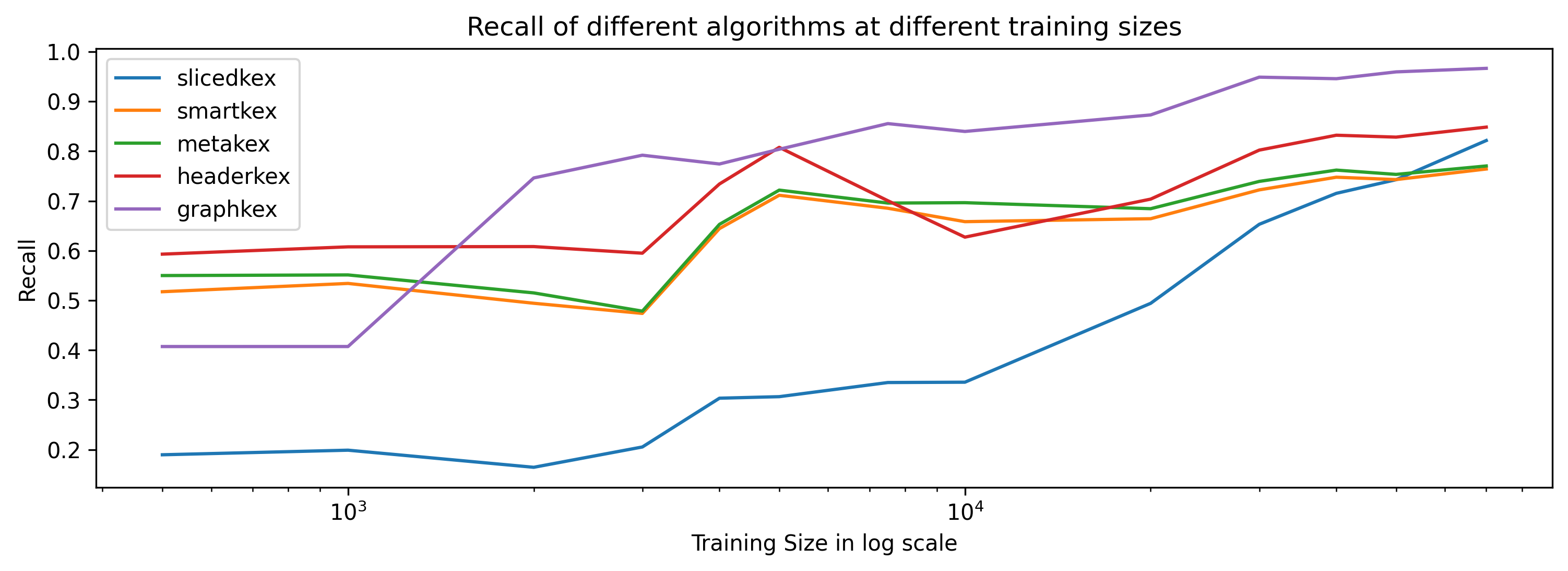}
    \caption{Recall of the different methods under different training data sizes. The x-axis is in log scale.}
    \label{machinekex_fig:recall}
\end{subfigure}
\hfill
\begin{subfigure}{0.45\textwidth}
    \includegraphics[width=\textwidth]{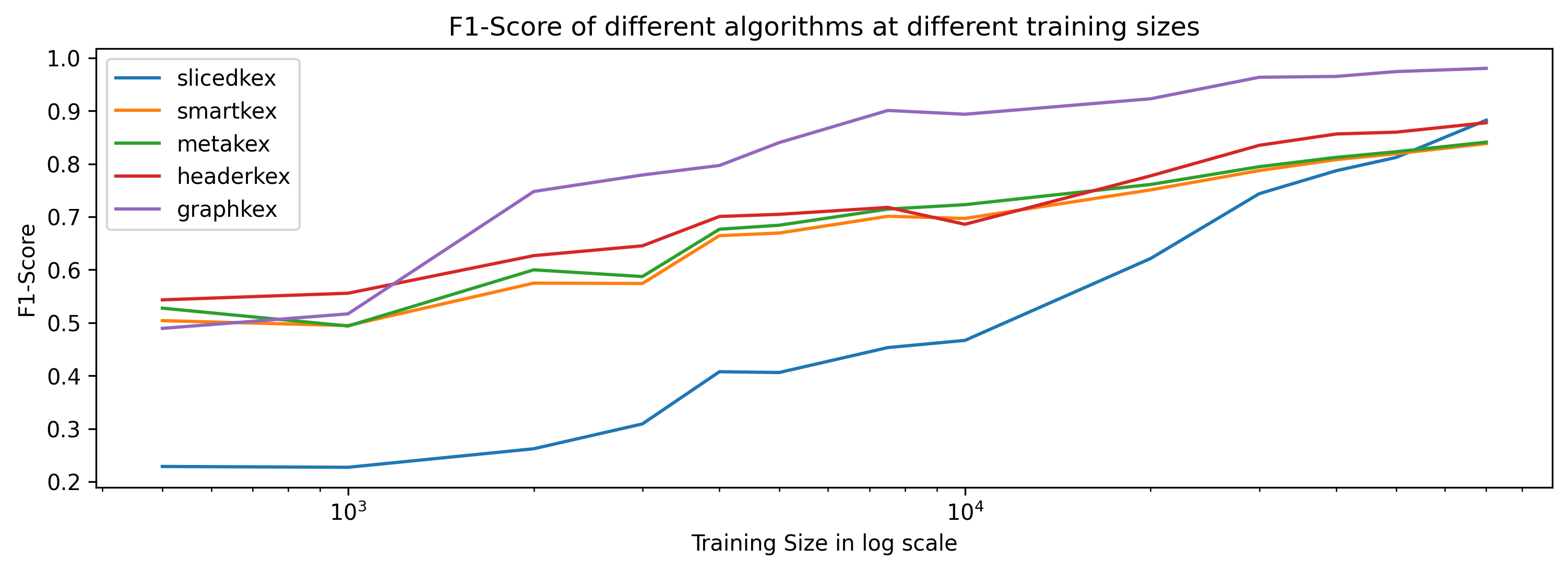}
    \caption{F1-Score of the different methods under different training data sizes. The x-axis is in log scale.}
    \label{machinekex_fig:f1}
\end{subfigure}
\hfill   
\caption{Plot of different metrics vs training instances}
\label{machinekex_fig:results_openssh}
\end{figure}

From the results in~\autoref{machinekex_fig:f1}, we see that the basic sliced approach, called~\slicedkex, lags behind all other methods.
With very little data present, approximately less than one thousand instances, this method performs the worst in comparison.
But when trained with a large amount of training data,~\slicedkex~performs better when compared to~\smartkex~and~\metakex.
This shows that having more training data helps the algorithm generalize better even without any help from feature engineering.
\headerkex~comes close to the performance of~\slicedkex~but with a significant reduction in training times and memory usage.
This could give an edge to the other algorithms simply due to their reduced usage of compute and memory.

In comparison, providing metadata provides way more information to the classifier to make a good prediction.
We see that our method MetaKex starts out with an F1-Score of over 45\%.
This by itself highlights the advantages of providing additional metadata to the classification algorithm.
As the training dataset size increases, all methods show notable improvements, but~\slicedkex~exhibits a particularly sharp performance boost once the number of heap instances exceeds 10,000, highlighting the impact of larger datasets on its effectiveness compared to the other methods.
The performance improvement w.r.t the number of training instances looks better when compared to the other methods as well.
\graphkex~performs the best due to having more information to identify the structures that we search.
But this information comes with specific assumptions about having a malloc header and address translation mechanisms, although translation of virtual addresses to physical addresses is one of the solved problems in the domain of Virtual Machine Introspection~(VMI).
Another assumption in this case is the structure of the data, and we know this because we are familiar with the source code of OpenSSH.
If OpenSSH releases a version with a different structure, the model may struggle to detect encryption keys. 
In such cases, we have two options: retrain the classifier with the new data or revert to a less metadata-dependent method.
There are a few dips in the performance during the initial stages, and we do expect that due to the randomized addition of data.
Towards the end, with enough data~\slicedkex~outperforms~\smartkex  which shows that more data could lead to better results.
But this increase in performance is hindered by the sheer amount of data we process and train.

When running for different methods, our notebook crashed due to the memory requirements on a server class machine with 1.5TB of memory.
We suspect it could be due to the combined memory requirements for all the algorithms when using the same parts of the dataset.
Therefore, we were not able to compute the final values for the training dataset size although those values were already computed for the different plots.

The final confusion matrix is given in~\autoref{machinekex_tab:ssh_confusion_matrix}.
Here, we see that the negative class vastly outnumbers all the others. 
The total instances for the methods based on~\smartkex~reduce the negative samples by more than 50\%. 
This accounts for the significant gain in training times for these methods.
\begin{table}[ht]
    \centering
    \begin{tabular}{|c|r|r|r|r|r|}
        \hline
        Method & True Negatives & True Positives & False Negatives & False Positives & Total Instances\\
        \hline
        SlicedKex &  99.4678 & 0.4308 & 0.0722 & 0.0291 & 21,934,385\\
        SmartKex & 99.0491 & 0.6662 & 0.1956 & 0.0891 & 9,805,636\\
        MetaKex & 99.0483 & 0.6756 & 0.1863 & 0.0899 & 9,805,636\\
        HeaderKex & 99.0368 & 0.7191 & 0.1427 & 0.1013 & 9,805,636\\
        GraphKex & 99.6375 & 0.3515 & 0.0068 & 0.0042 & 14,405,959\\
        \hline
    \end{tabular}
    \caption{Confusion matrix for the different methods as a percentage of the total instances. All values rounded to four decimal points.}
    \label{machinekex_tab:ssh_confusion_matrix}
\end{table}

We also compute the metrics for all the methods and we present it in~\autoref{machinekex_tab:ssh_metrics}.
Looking at~\autoref{machinekex_tab:test_cm}~we see that accuracy does not make any sense in this case as the dataset is very unbalanced.
It is better to consider the F1-Score of the methods in order to draw conclusions on the performance of each method.
From~\autoref{machinekex_tab:ssh_metrics}, we see that~\slicedkex~without any metadata gives impressive results.
~\slicedkex~improves over~\metakex~and~\headerkex~with more training data. 
Looking at~\autoref{machinekex_fig:f1}~,~\slicedkex~starts off being the worst algorithm when both~\metakex~and~\headerkex~are comparatively much better.
Another point to note is that although~\graphkex~has way more negative samples when compared to the other metadata based methods, the training time is very short due to the length of the engineered features.
\graphkex~is by far the best method among the others.
The training time for the algorithm is also the lowest among the others. 
But this involves more complex knowledge about the software data structures and requires virtual address to physical address translations.

\begin{table}[ht]
    \centering
    \begin{tabular}{|c|c|c|c|c|}
        \hline
        Method &  Accuracy & Precision & Recall & F1-Score\\
        \hline
        SlicedKex & 99.89 & 93.67 & 85.63 & 89.48\\
        SmartKex & 99.72 & 88.20 & 77.29 & 82.39\\
        MetaKex & 99.72 & 88.26 & 78.38 & 83.03\\
        HeaderKex & 99.75 & 87.65 & 83.34 & 85.49\\
        GraphKex & 99.98 & 98.82 & 98.10 & 98.46\\
        \hline
    \end{tabular}
    \caption{Metrics for different methods.}
    \label{machinekex_tab:ssh_metrics}
\end{table}
\begin{figure}
\centering
\begin{subfigure}{0.45\textwidth}
    \includegraphics[width=\textwidth]{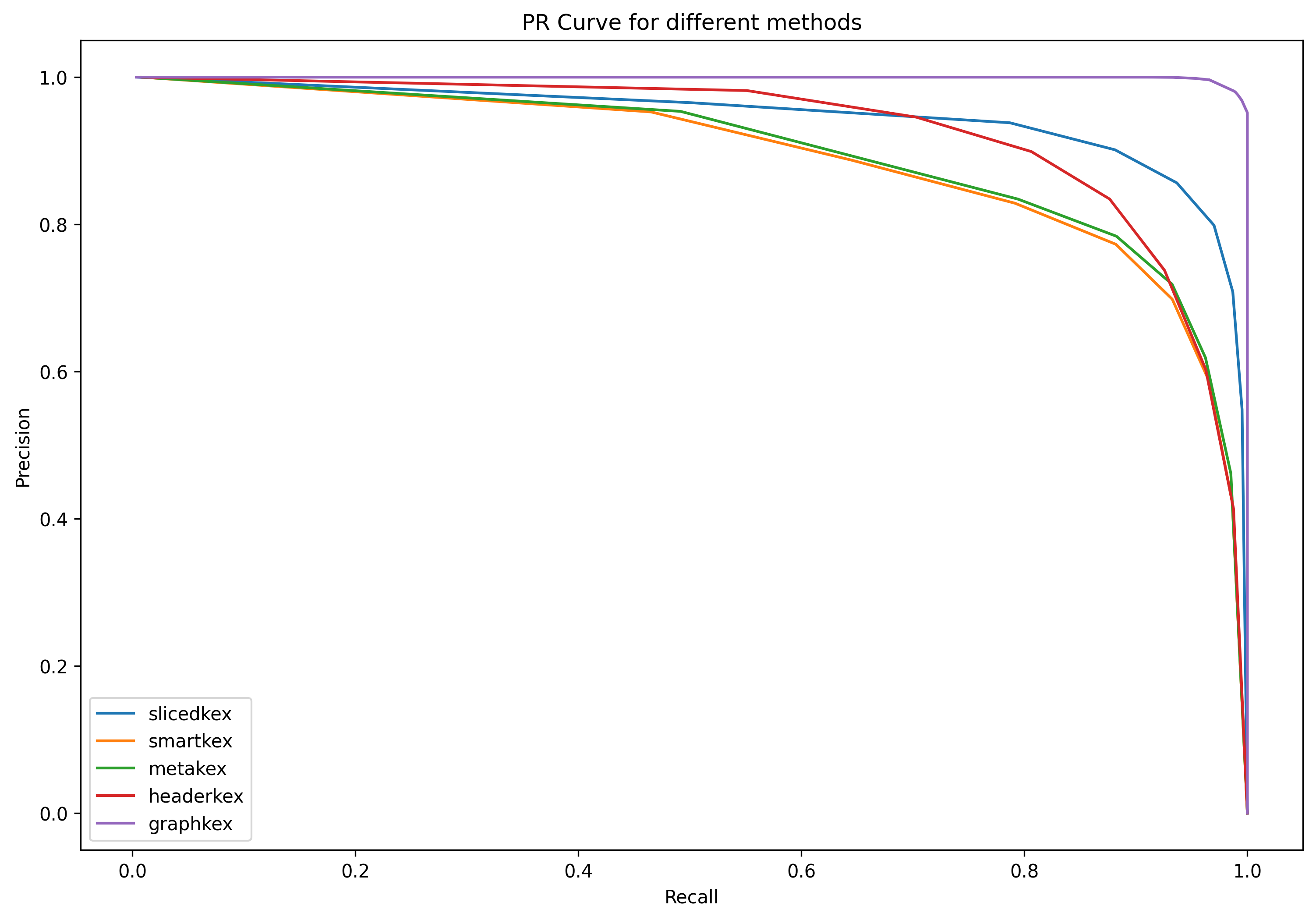}
    \caption{Precision-Recall curve of different methods.}
    \label{machinekex_fig:ssh_pr}
\end{subfigure}
\hfill
\begin{subfigure}{0.45\textwidth}
    \includegraphics[width=\textwidth]{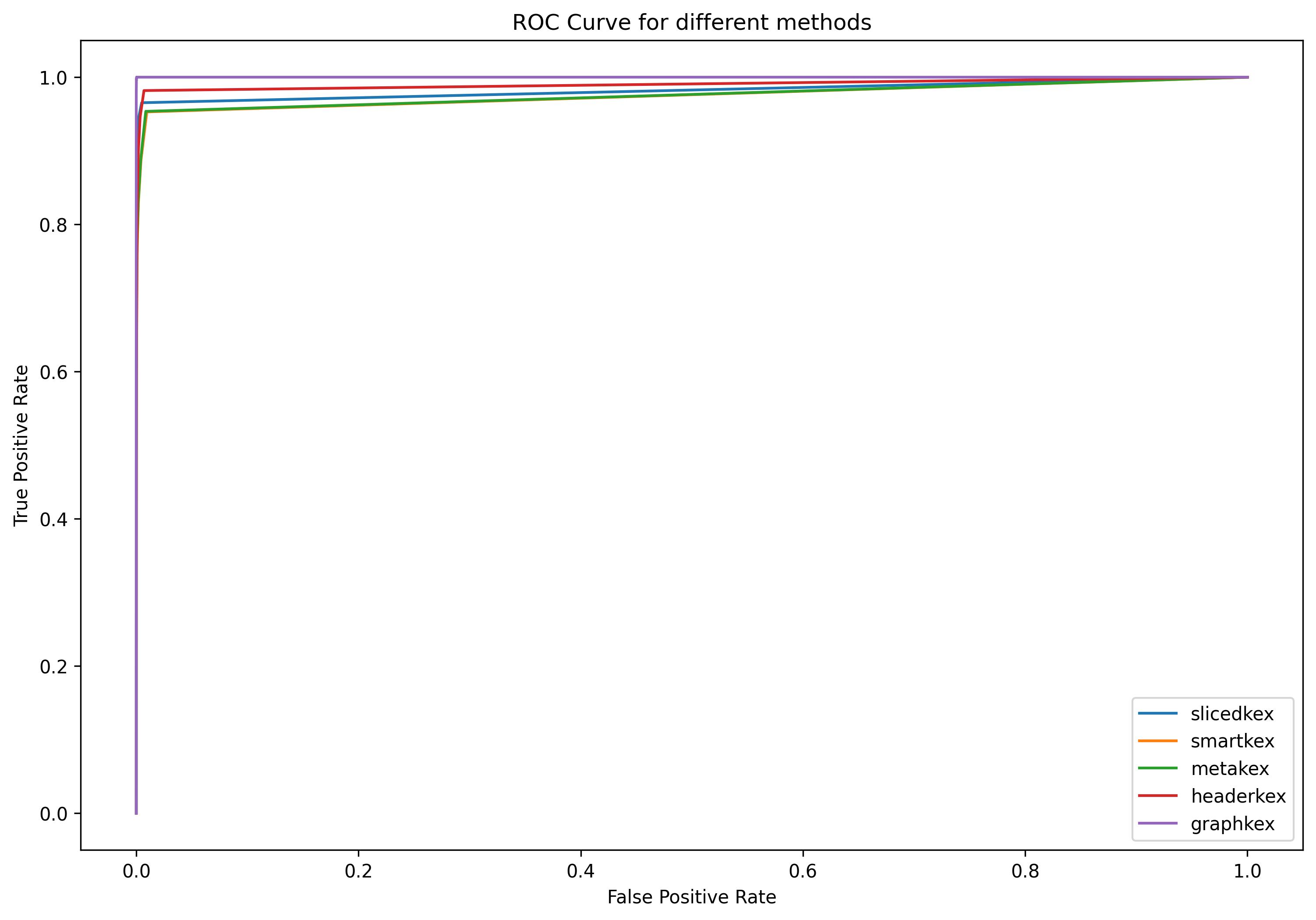}
    \caption{ROC Curve of different methods.}
    \label{machinekex_fig:ssh_roc}
\end{subfigure}
\hfill   
\caption{Plots of Precision-Recall and ROC Curves using the complete training set for training and tested on the validation dataset.}
\label{machinekex_fig:ssh_curves}
\end{figure}

\subsection{Main Memory Snapshots}
For the classification of main memory snapshots, we employ a two-step approach. 
The first step is to identify whether a given pointer is of interest, meaning it points to a data structure relevant in the context of Virtual Machine Introspection (VMI). 
In the second step, the identified pointer is classified into its associated data structure. 
This two-step classification is necessary because the physical memory contains thousands of pointers, and performing multiclass classification directly would reduce efficacy due to the high proportion of non-relevant pointers.
More than~98\%~of the pointers found in the main memory snapshot are of no interest to us in this cope.
The reference of these pointers are also diverse in size and structure, leading to complex decision boundaries. 
This diverse nature of results in deep decision trees and diminished performance.
Therefore, we found it better to have a two-stage classification process and this leads to smaller sized classifiers and better performance.

Our dataset has over two thousand main memory snapshots but we use fifty-two of them to train the random forest.
We then create the feature vector dataset from these fifty snapshots and do a 25\% train-test split on this dataset.
This is our validation dataset and the test dataset consists of fifty completely different snapshots that are not seen by the classifiers during both training and validation testing.
In this case, our training dataset is smaller than the testing dataset due to splitting of the training dataset into training and validation.
There are also three runs of the experiment with three different training and testing subsets.
This allows us to better compare the capability of methods and avoid lucky cherry picking for the training test splits.

We present our results for the validation and testing splits in~\autoref{machinekex_tab:binary_classification}~and~\autoref{machinekex_tab:multiclass_classification}.
\begin{table}[ht]
    \centering
    \begin{tabular}{|c|c|c|}
         \hline
         Metric &  Validation Set & Test Set\\
         \hline
         Accuracy & 99.84 & 99.83\\
         Precision & 96.91 & 95.23\\
         Recall & 91.14 & 89.51\\
         F1-Score & 93.94 & 92.28\\
         \hline
    \end{tabular}
    \caption{Results on the validation and test splits for identifying whether a pointer is of interest or not. This is a binary classification problem.}
    \label{machinekex_tab:binary_classification}
\end{table}
\begin{table}[ht]
    \centering
    \begin{tabular}{|c|c|c|}
         \hline
         Metric &  Validation Set & Test Set\\
         \hline
         Accuracy & 99.84 & 99.85\\
         Precision & 97.70 & 98.47\\
         Recall & 95.18 & 93.69\\
         F1-Score & 95.91 & 94.77\\
         \hline
    \end{tabular}
    \caption{Results on the validation and test splits for identifying the structure using the corresponding pointer. This is a multiclass classification problem.}
    \label{machinekex_tab:multiclass_classification}
\end{table}
Looking at the confusion matrices for both the validation split and test split in~\autoref{machinekex_tab:validation_cm}~and~\autoref{machinekex_tab:test_cm}, we see how the negative class makes up most of the dataset.
Even for such a highly unbalanced dataset, the results are very precise. This shows that our classifier performs very well and is discarding unwanted data.
\begin{table}[ht]
    \centering
    \begin{tabular}{|c|r|r|}
         \hline
           & $y_{pred}=0$ & $y_{pred}=1$ \\
         \hline
         $y_{true}=0$ & 14,984,882 & 6,047 \\
         $y_{true}=1$ & 18,453 & 189,809 \\
         \hline
    \end{tabular}
    \caption{Confusion matrix for binary classification on the validation split.}
    \label{machinekex_tab:validation_cm}
\end{table}
\begin{table}[ht]
    \centering
    \begin{tabular}{|c|r|r|}
         \hline
           & $y_{pred}=0$ & $y_{pred}=1$ \\
         \hline
         $y_{true}=0$ & 36,279,816 & 17,688 \\
         $y_{true}=1$ & 41,344 & 352,898 \\
         \hline
    \end{tabular}
    \caption{Confusion matrix for binary classification on the test split.}
    \label{machinekex_tab:test_cm}
\end{table}

\begin{figure}
\centering
\begin{subfigure}{0.49\textwidth}
    \includegraphics[width=\textwidth]{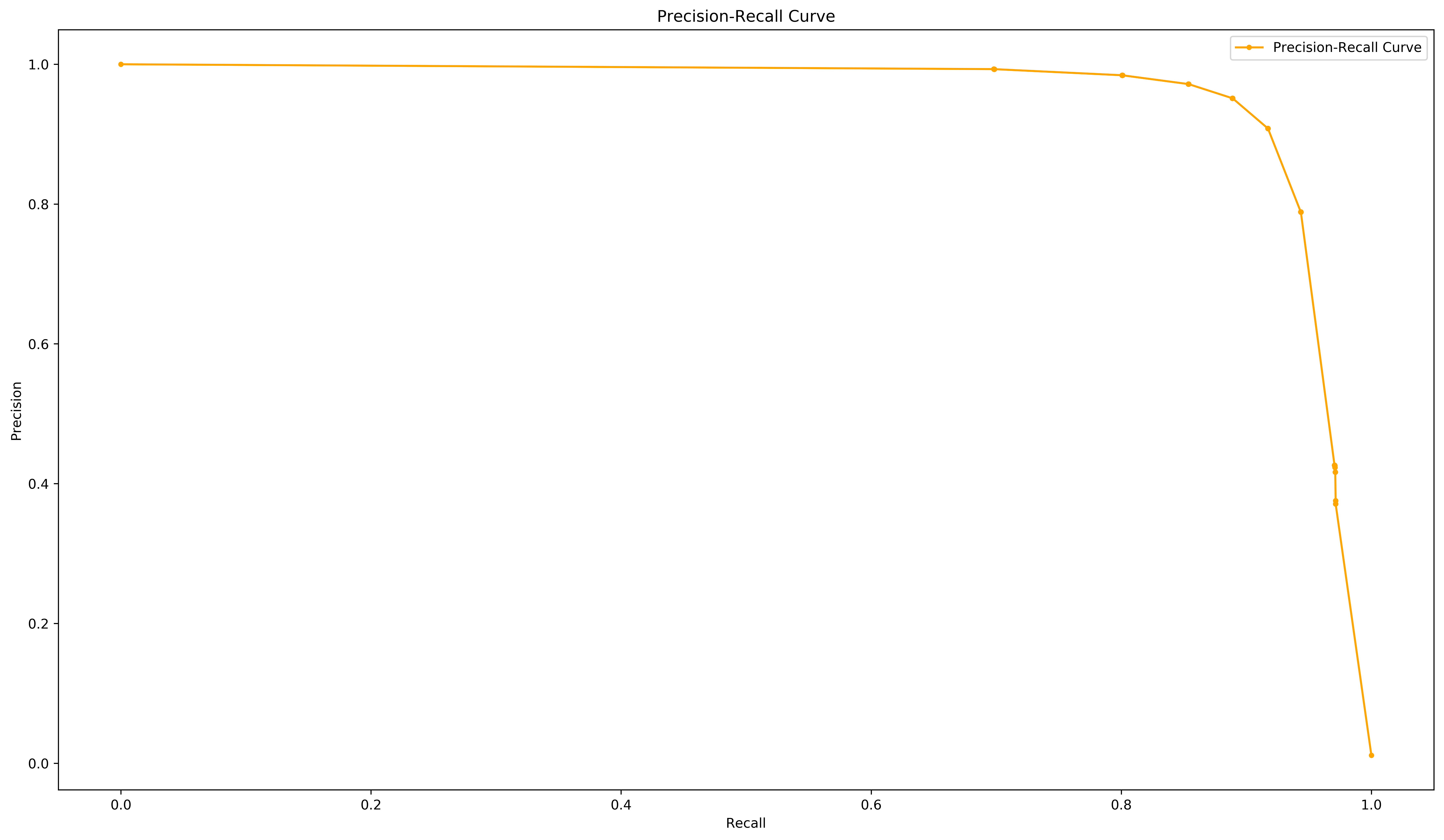}
    \caption{Precision Recall Curve for memory snapshots.}
    \label{machinekex_fig:mem_snapshot_pr}
\end{subfigure}
\hfill
\begin{subfigure}{0.49\textwidth}
    \includegraphics[width=\textwidth]{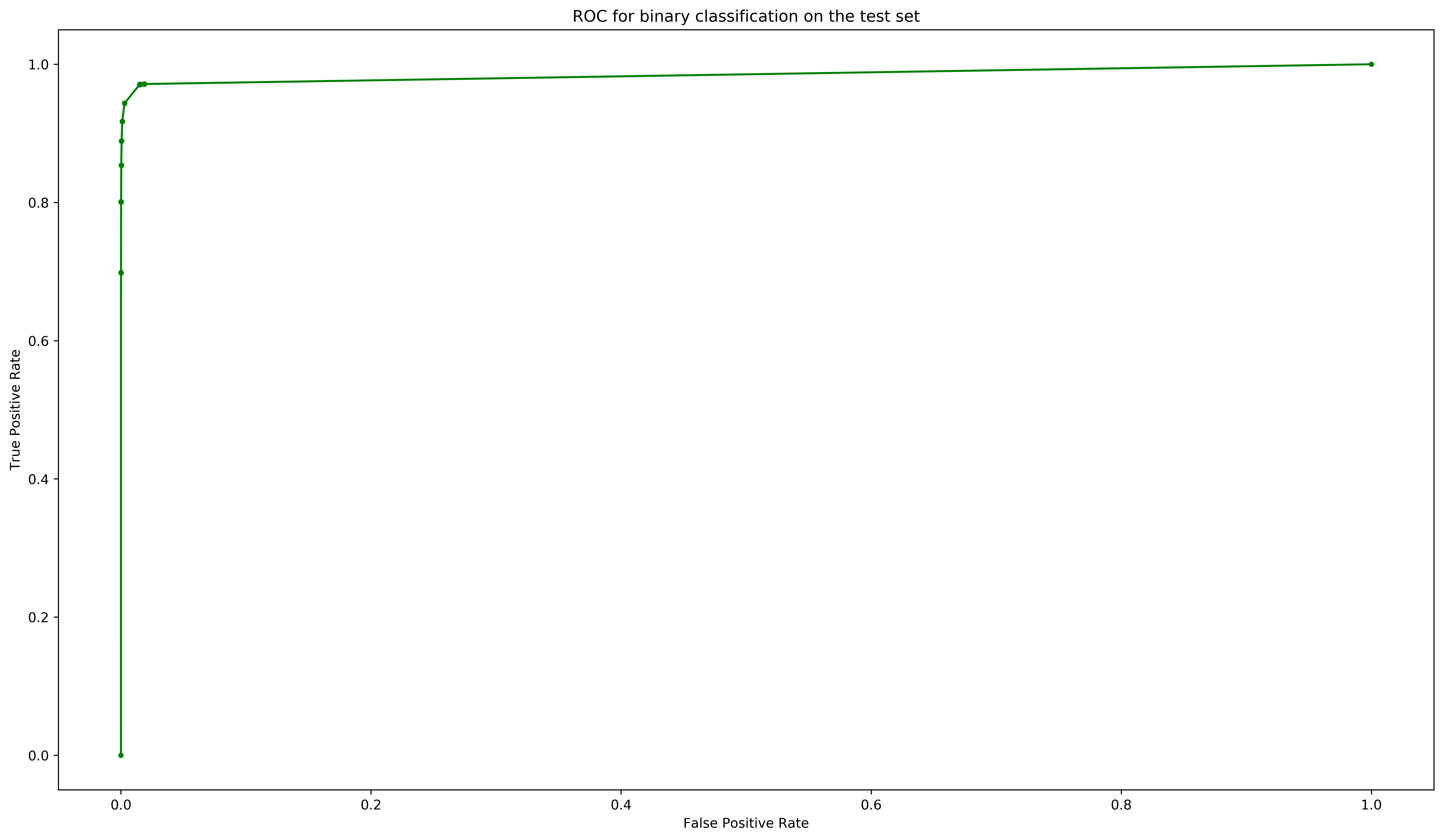}
    \caption{ROC Curve for memory snapshots.}
    \label{machinekex_fig:mem_snapshot_roc}
\end{subfigure}
\hfill
\caption{Plot of Precision-Recall and ROC curves for memory snapshots.}
\label{machinekex_fig:mem_snapshot_curves}
\end{figure}
We observe that even with equal or less amounts of training data, our classifier performs very well. 
While there is a drop in recall performance when compared to precision performance, overall the method works exceptionally well on the test set.

The Area Under Curve~(AUC)~for each of the runs of the memory snapshot is in~\autoref{machinekex_tab:mem_snapshot_auc}.
Our model performs excellently on discriminating between relevant pointers to non-relevant pointers identified within the memory snapshot.
\begin{table}[ht]
    \centering
    \begin{tabular}{|c|c|}
    \hline
    Run  & Value \\
    \hline
    Run 1 & 0.9537 \\
    Run 2 & 0.9537 \\
    Run 3 & 0.9537 \\
    \hline
    \end{tabular}
    \caption{Area under the curve for each run}
    \label{machinekex_tab:mem_snapshot_auc}
\end{table}

\section{Ethical Considerations}
\label{machinekex_sec:ethical_considerations}
Developing and applying techniques for detecting and extracting keys from memory underscore crucial ethical considerations. 
Paramount among these is respect for privacy and data protection, which mandates strict compliance with pertinent laws and guidelines. 
Informed consent from participants and third parties, where applicable, is imperative, alongside adherence to responsible disclosure practices to mitigate potential risks. 
It is essential to emphasize that the method and its corresponding code are solely intended for research purposes and must not be utilized for malicious activities.
\section{Conclusion}
\label{machinekex_sec:conclusion}
In this study, we analysed how metadata could enhance the performance of classifiers in the context of Virtual Machine Introspection.
Our methods demonstrate impressive performance even with limited training data. 
Despite the data imbalance, we obtained an F1-Score better than~80\%.
Our work underlines the importance of leveraging metadata for feature engineering. 
The results demonstrate that supervised techniques can bridge the semantic gap present in VMI and FMA, providing valuable assistance to cybersecurity experts.
Additionally, our experiments also emphasise the importance of feature engineering and a priori knowledge for improving the results over standard algorithms.
We found that the more information we extract from the raw data, the better the results, particularly when working with limited training data.
As a use case, we applied our findings into identifying memory structures in kernel data from memory snapshots.
Our feature engineering and learning methods excel in both identifying relevant pointers and classifying those pointers into different types.
This shows that our method could help memory forensic analysts and advance the field of Virtual Machine Introspection.

\bibliography{bibliography/machinekex}
\bibliographystyle{bibliography/tmlr}

\end{document}